
\parskip=12pt
\baselineskip=18pt

\def\CC#1={\vskip\the\parskip\centerline{\bf#1}}
\def\cc#1={\centerline{#1}}
\def\HALF{{\textstyle{1\over2}}}
\def\ST{\sqrt{3}}

	~\vskip 24pt
\CC Evolutionary Laws, Initial Conditions, and Gauge Fixing=
\CC in Constrained Systems=
	\vskip 12 pt

	\vskip 12 pt
\cc J M Pons\dag$^1$ and L C Shepley\ddag=
	\vskip 24 pt
		\cc \dag Dep.\ d'Estructura i Constituents de la Mat\`eria, Universitat
de Barcelona,=
		\cc Av.\ Diagonal 647, E-08028 Barcelona, Catalonia, Spain=
	\vskip 12pt

\cc\ddag Physics Department, The University of Texas, Austin, Texas 78712, USA=

\vskip 24pt

{\bf{}Abstract}.
We describe in detail how to eliminate nonphysical degrees of freedom
in the Lagrangian and Hamiltonian formulations of a constrained
system.  Two important and distinct steps in our method are the fixing
of ambiguities in the dynamics and the determination of inequivalent
initial data.  The Lagrangian discussion is novel, and a proof is given
that the final number of degrees of freedom in the two formulations
agrees.  We give applications to reparameterization invariant theories,
where we prove that one of the constraints must be explicitly time
dependent.  We illustrate our procedure with the examples of
trajectories in spacetime and with spatially homogeneous cosmological
models.  Finally, we comment briefly on Dirac's extended Hamiltonian
technique.

\vskip 24 pt

Short Title:  Evolutionary Laws and Gauge Fixing
\vskip 24pt


\vfill
\hrule

$^1$Supported by the Comisi\'on Interministerial para
la Ciencia y la Tecnolog\'\i{}a (project number AEN-0695) and by a
Human Capital and Mobility Grant (ERB4050PL930544)

\vfill\eject

\CC 1 Introduction=

Dynamical theories exhibiting gauge freedom are described by singular
Lagrangians or in the Hamiltonian formalism
introduced by Dirac [1,2].  The existence of constraints
reduces the true degrees of freedom in the system.  This reduction
has specific features depending upon whether we are in velocity space
(Lagrangian formalism) or in
phase space (Hamiltonian formalism).  In particular the dimensions of
the constraint surfaces are
different:  One can prove [3] that the
number of constraints in the Hamiltonian formalism equals the number of
constraints in the Lagrangian formalism plus the number of independent
gauge transformations.  In other words, the number of degrees of
freedom in phase space seems to be less than the number in velocity
space, but getting rid of the gauge freedom in the phase and velocity
spaces eliminate spurious degrees of freedom.  We will
show that therefore the number
of degrees of freedom are the same in the two formalisms.

There are two different stages in the reduction of the
degrees of freedom. The first is provided by the constraints that
naturally arise in the formalism by pure consistency requirements. The second
corresponds to the gauge fixing procedure.  In our case it will consist
in the introduction of new constraints in such a
way that the gauge degrees of freedom are thoroughly eliminated.

In [4] an analysis was carried out on how to implement a gauge
fixing procedure in the Hamiltonian formalism to obtain the true number of
physical degrees of freedom.
In the present paper we will proceed to general Lagrangian
gauge fixing in a way
completely independent from the Hamiltonian method.

We will emphasize the double role played by the gauge fixing
procedure. Some constraints are needed to fix the time
evolution of the gauge system, which was undetermined to a certain extent. The
rest of the constraints eliminate the spurious degrees of freedom that are
still
present in the setting of the initial conditions of the system. This double
role may be
relevant for quantization (we treat only classical systems here).
It has not been adequately emphasized in the literature.

The gauge fixing procedures in the Hamiltonian and Lagrangian
formalisms are technically different.  For instance, there is no
Poisson Bracket available in velocity space for singular Lagrangians,
and the relationship between velocity and position in the tangent
bundle does not carry over to the cotangent bundle.  This difference is
why there are difficulties in implementing a Lagrangian gauge fixing
procedure which is independent of a pullback of the Hamiltonian one.
We here show how to fix these difficulties and how to perform in full
generality gauge fixing in both formalisms.  Hence our results will include
a general proof of the matching of the degrees of freedom in Hamiltonian
and Lagrangian formalisms for gauge theories.

Our paper is organized as follows: In section 2 we develop a detailed
version of the Hamiltonian gauge fixing procedure (with some
improvements to [4]). In section 3 we study the
Lagrangian gauge fixing procedure and establish
the theorem that the number of degrees of freedom in the Hamiltonian
and Lagrangian formalisms are equal.
In section 4 we describe some details concerning reparameterization
invariant theories, since in section 5 we apply our results to two
such theories. These examples are the parameter-independent Lagrangian
for geodesics in special relativity and the Types I and IX spatially
homogeneous
cosmologies, and we will emphasize the role of time-dependent gauge fixing
conditions.

Throughout the paper we will assume that some regularity conditions are
fulfilled: The Hessian matrix of the Lagrangian with respect to
the velocities has constant rank,
ineffective constraints (such that their gradient vanishes on the
constraint surface) do not appear,
and also the rank of the Poisson Bracket
matrix of constraints remains constant in the stabilization algorithm
(so that a second class constraint can never become first class
by adding new constraints to the theory).

We emphasize that we always maintain the equivalence
between the Lagrangian and the Hamiltonian formalisms [5].
This equivalence holds even before the implementation of the
gauge fixing conditions; in particular we do not modify the
Hamiltonian formalism by adding {\it{}ad hoc\/} constraint
terms as Dirac has proposed [2,6,7].
This proposal has been proved
to be unnecessary under our regularity conditions [4].
We will discuss this point more fully in the Conclusion.

\CC 2 Hamiltonian Gauge Fixing Procedure=

We start with a canonical formalism using Dirac's method, starting from
a singular Lagrangian $L(q^i,\dot q^i)$ $(i=1,\cdots,N)$
($\dot q^i=dq^i/dt$).  The functions
$\hat p_i(q,\dot q)=\partial L/\partial\dot q^i$ are used to define
the Hessian $W_{ij}=\partial\hat p_i/\partial\dot q_j$, a matrix with rank
$N-P$ (we assume this rank is constant), $P$ being the number of
primary constraints.  The Legendre map from
velocity space (tangent bundle for configuration space) $TQ$ to phase
space (cotangent bundle) $T^*\!Q$ defined by $p_i=\hat p_i(q,\dot q)$
defines a constraint surface of dimension $2N-P$.

The function $E_L:=\hat p_i\dot q^i-L$ in velocity space
(the so-called energy function) is mapped to a
function on the constraint surface, and in
phase space a canonical Hamiltonian $H_C$ may
be defined which agrees with this function on the surface.  $H_C$ is not
unique, and to it may be added a linear combination
$\lambda^\rho\psi_\rho(q,p)$ of constraint functions $\psi_\rho
(\rho=1,\cdots,P)$, the $\lambda^\rho$ being arbitrary functions of
time $t$ and the vanishing of $\psi_\rho(q,p)$ defining the primary
constraint surface.
These primary constraints may be chosen so that some are first class
(their Poisson Brackets with all the constraints weakly vanish,
that is, vanish on the constraint surface) and some
are second class (the matrix of their Poisson Brackets with each
other is nonsingular---there must be an even number of second
class constraints or none).

Let $\psi^1_\rho,\psi^2_\rho$ denote first and second class primary
constraints, $\lambda^\rho_{1,2}$ being the respective $\lambda$
functions.  The time
derivatives of the second class constraints yield
	$$	\dot\psi^2_\rho=\{\psi^2_\rho,H_C\}
		+ \lambda_2^{\sigma}\{\psi^2_\rho,\psi^2_\sigma\}\ ,
	$$
equations which may be solved for the $\lambda_2^\sigma$ by requiring
$\dot\psi^2_\rho=0$.  These functions $\lambda_2^\sigma$ are then
inserted into the expression $H_C +\lambda_1^\rho\psi^1_\rho
+\lambda_2^\rho\psi^2_\rho$ to yield a new candidate for
the Hamiltonian:
	$$	H_C +\lambda_1^\rho\psi^1_\rho +\lambda_2^\rho\psi^2_\rho
			=H^1_C +\lambda_1^\rho\psi^1_\rho\ .
	$$

The time derivatives of the primary first class constraints yield
	$$	\dot\psi^1_\rho=\{\psi^1_\rho,H_C\}\ ,
	$$
which can either be zero on the constraint surface or not.
In the latter case, new
constraints are found.  The time derivatives of these new constraints
will involve their Poisson Brackets with $H^1_C$ and with the $\psi^1_\rho$
(and will also involve their partial time derivatives).
The requirement that these time derivatives vanish will be equations,
some of which can be solved for some of the $\lambda_1^\rho$
functions.  The rest of these equations may yield more constraints,
and the process of requiring the vanishing of their time derivatives
is repeated at a deeper level.

This brief description of the stabilization algorithm is not meant to
be rigorous, but the process does eventually finish.  Once the stabilization
algorithm has been performed, we end up with [2,5,8]:

\item{1:}
	A certain number, $M$, of constraints.  These constraints may
	be more numerous than the ones introduced above ($M\geq P$),
	but they are arranged into first and second
	classes.  The first class constraints have weakly vanishing
	Poisson Brackets with all the constraints, and the matrix of
	the second class constraint Poisson Brackets is nonsingular.
	These constraints restrict the dynamics to a
	constraint surface within $T^*\!Q$ of dimension $2N-M$.

\item{2:}
	A dynamics (with some gauge arbitrariness) on the constaint surface
	which is generated,
	through Poisson Brackets, by the so called Dirac Hamiltonian:
	$$	H_D:= H_{FC} + \lambda^{\mu}\phi^1_{\mu}\ .
	$$
	$H_{FC}$ is the first class Hamiltonian, obtained by
	adding to the canonical Hamiltonian $H_C$ pieces linear in the
	primary second
	class constraints.  $\phi^1_{\mu}$ $(\mu = 1,\cdots,P_1)$ are the
	primary (hence the superscript $1$) first class constraints.
	The secondary and higher first class constraints, obtained from
	the time derivatives of the $\phi^1_{\mu}$, are not used here.
	The $\lambda^{\mu}$ are arbitrary functions of time (or
	spacetime in field theories).

\item{3:}
	A certain number ($P_1$) of independent gauge transformations
	generated, through Poisson Brackets,
	$$	\delta_\mu q^i=\{q^i,G_\mu\}\ ,\ \delta_\mu p_i=\{p_i,G_\mu\}\ ,
	$$
	by functions
	$G_{\mu}$ $(\mu=1,\cdots,P_1)$ which have the following form
	[4,9,10,11]:
	 $$	G_{\mu} = \epsilon_{\mu} \phi^{K_{\mu}}_{\mu}
			+ \epsilon^{(1)}_{\mu} \phi^{K_{\mu}-1}_{\mu}
			+ \epsilon^{(2)}_{\mu} \phi^{K_{\mu}-2}_{\mu}
			+ \cdots
			+ \epsilon^{(K_{\mu}-1)}_{\mu} \phi^1_{\mu}\ ,
	$$
	where $\epsilon_{\mu}$ is an arbitrary infinitesimal function
	of time; $\epsilon^{(r)}_{\mu}$ is its r-th time derivative;
	$K_{\mu}$ is the length of the stabilization algorithm for the
	primary  first class constraint $\phi^1_{\mu}$; and
	$\phi^2_{\mu},\dots,\phi^{K_{\mu}}_{\mu}$ are secondary through
	$K_\mu$-ary, first class constraints.  It turns out [12] that
	one can take these gauge generators in such a way that all the
	first class constraints are involved once and only once in the
	$G_\mu$, and so their total number equals
	$$	F:=\sum^{P_1}_{\mu=1}K_\mu\ .
	$$

Now we are ready for the gauge fixing procedure.
Even though the order of introducing the
gauge fixing constraints is irrelevant,
we will proceed in the way that makes the whole
procedure more illuminating from the theoretical point of view.
As we said in the Introduction, we will distinguish two different
steps in the gauge fixing procedure, corresponding to
evolutionary laws and initial conditions [4].  In the first step we fix the
laws of evolution, which otherwise have
a certain amount of mathematical arbitrariness. In the second step we
eliminate the redundancy of initial conditions that are physically
equivalent.

The arbitrariness in the
dynamics is represented by the $P_1$ functions $\lambda^{\mu}$.
To get rid of this arbitrariness, we introduce
a set of $P_1$ constraints $\chi^1_{\mu}\simeq0$ $(\mu = 1,\cdots, P_1)$,
defined so that their own
stability equations, under dynamical evolution, will determine the functions
$\lambda^{\mu}$.  To this end we must require that the matrix
	$$	C_{\mu\nu}:= \{ \chi^1_{\mu}, \phi^1_{\nu} \}
	$$
be non-singular. The conservation in time of this new set of
constraints leads to
	$$	\dot\chi^1_\mu=0 = {\partial\chi^1_\mu\over\partial t}
			+ \{\chi^1_{\mu},H_{FC}\}
			+ \lambda^{\nu}\{\chi^1_{\mu},\phi^1_{\nu}\}\ ,
	$$
which determines $\lambda^{\nu}$ as
	$$	\lambda^{\nu} = -(C^{-1})^{\nu\mu}
			\bigr(\{\chi^1_{\mu},H_{FC}\}
				+{\partial\chi^1_\mu\over\partial t}\bigl)\ .
	$$
The dynamical evolution thus becomes completely determined.
The imposition of these constraints causes the dynamics to be further
restricted to the ($2N-M-P_1$)-dimensional constraint surface defined by
$\chi^1_{\mu}=0$.

The gauge fixing procedure is not yet finished.
It is necessary to address the issue of initial conditions, which
we call ``point gauge equivalence,'' our second step.  Let us clarify
this crucial point:
Even though the dynamics has now been fixed, there is still the
possibility of gauge transformations which take one trajectory into
another.  To check whether these gauge transformations do exist, we
need only check their action at a specified time.  That is, the points
on the set of trajectories at a specified time are unique initial data
for the trajectories.  If a gauge transformation exists which relates
two initial data points, then these two points are physically
equivalent.
We will
obtain the generators of the transformations which take initial points
into equivalent ones (``point gauge transformations'') and use them
to fix the gauge finally.

Consider the gauge generators given above at, say for simplicity,
$t=0$: $G_{\mu}(0)$. The
most arbitrary point gauge transformation at $t=0$ will be generated by
$G(0)=\sum_{\mu=1}^{P_1}G_{\mu}(0)$.
The arbitrary functions $\epsilon_{\mu}$ and their derivatives
become, at the given time $t=0$,
(infinitesimal) independent arbitrary parameters (there are F in
number).  We redefine them as
	$$	\alpha_{\mu,i_{\mu}}:=\epsilon^{(K_{\mu}-i_{\mu})}_{\mu}(0)\ ;\
			(\mu = 1,\cdots,P_1)\ (i_{\mu}=1,\cdots,K_{\mu})\ .
	$$

These point gauge transformation generators must be consistent with the
new constraints $\chi^1_{\mu}$.
This requirement introduces relations among the $\alpha_{\mu,i_{\mu}}$:
	$$	0 = \{\chi^1_{\nu},G(0)\}
		=\{\chi^1_{\nu}, \sum_{\mu=1}^{P_1}\sum_{i_{\mu}=1}^{K_{\mu}}
			\alpha_{\mu,i_{\mu}} \phi^{i_{\mu}}_{\mu}\}
		=\sum_{\mu=1}^{P_1}\Bigl(\sum_{i_{\mu}=2}^{K_{\mu}}
			\alpha_{\mu,i_{\mu}} \{\chi^1_{\nu},\phi^{i_{\mu}}_{\mu}\}
			+ \alpha_{\mu,1} C_{\nu\mu}\Bigr)\ .
	$$
Remember that the matrix $C_{\nu\mu}=\{\chi^1_\nu,\phi^1_\mu\}$ is
nonsingular.  These relations imply
	$$	\alpha_{\rho,1}
		= -(C^{-1})^{\rho\nu}\sum_{\mu=1}^{P_1}\sum_{i_{\mu}=2}^{K_{\mu}}
			\alpha_{\mu,i_\mu} \{\chi^1_{\nu},\phi^{i_{\mu}}_{\mu}\}\ .
	$$
As a consequence, the independent point gauge generators are
	$$	\tilde\phi^{i_\mu}_{\mu} :=
			\phi^{i_\mu}_\mu - \phi^1_\rho (C^{-1})^{\rho\nu}
			\{\chi^1_\nu,\phi^{i_\mu}_\mu\}\
			(\mu=1,\cdots,P_1)\ (i_\mu=2,\cdots,K_{\mu})\ .
	$$
Notice that new point gauge generators only exist when there are secondary
first-class constraints, that is, when the length $K_\mu$ of at least one of
the $G_\mu$ is greater than one.

Recall that $F$ is the number of first class constraints in the original
theory, including primary, secondary, and higher constraints; we
conclude that there are $F-P_1$ generators that relate physically equivalent
initial conditions.  To eliminate the extraneous variables,
we will select a unique representative of each equivalence class by
introducing a new set of $F-P_1$ gauge fixing constraints,
$\chi_{\mu}^{i_{\mu}} \simeq 0$ $(\mu=1,\cdots,P_1)$
$(i_{\mu}=2,\cdots,K_{\mu})$, such that
	$$	\det|\{\chi_\mu^{i_\mu},\tilde\phi_{\nu}^{j_\nu}\}| \neq 0,\
			i_\mu\neq 1,\ j_\nu\neq 1\ ,
	$$
in order to prevent any motion generated by $\tilde\phi_\nu^{j_\nu}$.
The stability requirement is
	$$	{\partial\over\partial t}\chi_\mu^{i_\mu} +
		\{\chi_\mu^{i_\mu},H_D\} \simeq 0\ ,
	$$
which $\simeq0$ means vanishing on the constraint surface; this
requirement simply dictates how the $\chi_\mu^{i_\mu}$ evolve off the
initial data surface.

Notice that we have explicitly allowed time
dependence in the $\chi_\mu^{i_\mu}$ constraints.  In fact,
time dependence is necessary in the special case when $H_{FC}$
is a constraint (first class, of course).  This point will
be clarified in Section 4.

This ends the gauge fixing procedure. Now we can count
the physical number of degrees of
freedom:  The $M$ constraints left after the stabilization algorithm
restricted motion to a $2N-M$-dimensional surface in $T^*\!Q$.  The gauge
fixing constraints needed to fix the evolutionary equations number
$P_1$.  Finally there are $F-P_1$ point gauge fixing constraints needed to
select physically inequivalent initial points.  (The total number of
gauge fixing constraints equals the number $F$ of first class
constraints in the original theory.)  The final number of degrees of freedom
is $2N-M-F$.
Notice that $M-F$ is the original number of second class constraints
and is therefore even.  Consequently, $2N-M-F$ is even; this result
agrees with the fact that the above procedure makes all constraints
into second class ones, and in this case the constraint surface is
symplectic [13].

\CC 3 Lagrangian Gauge Fixing Procedure=

We first use a stabilization algorithm similar to the one used in the
Hamiltonian formalism.  (In the equations below, we use the summation
convention for configuration space indices $i=1,\cdots,N$.)  The
equations of motion obtained from the Lagrangian $L$ are (assuming for
simplicity no explicit time dependence):
$$	W_{is}\ddot q^s = \alpha_i\ ,
$$
where
$$	W_{ij}= {\partial^2L\over\partial\dot q^i\partial\dot q^j}\ ,\
	\alpha_i= - {\partial^2L\over\partial\dot q^i\partial q^s}\dot q^s
		+ {\partial L\over\partial q^i}\ .
$$
If $W_{ij}$ is singular, it possesses $P$ null vectors $\gamma^i_\rho$,
giving up to $P$ (these relations may not be independent) constraints
$$	\alpha_i \gamma^i_\rho\simeq 0\ .
$$
It is easily shown that there exists at least one $M^{ij}$ and
$\tilde\gamma^\rho_i$ such that
$$	W_{is}M^{sj}= \delta^j_i + \tilde\gamma^\rho_i \gamma^j_\rho\ ,
$$
and therefore [14]
$$	\ddot q^i= M^{is}\alpha_s+ \tilde\eta^\rho\gamma^i_\rho\
		({\rm{}with\ }\tilde\eta^\rho= \tilde\gamma^\rho_i\ddot q^i)\ ,
$$
where $\tilde\eta^\rho$ are arbitrary functions of $t$.

The stabilization algorithm starts by demanding that time evolution
preserve the $\alpha_i\gamma^i_\rho$ constraints.
Sometimes new constraints are found;
sometimes some of the $\tilde\eta^\rho$ are determined; eventually the
dynamics is described by a vector field in velocity space
$$	X:= {\partial\over\partial t}
		+ \dot q^i {\partial\over\partial q^i}
		+ a^i(q,\dot q){\partial\over\partial\dot q^i}
		+\eta^\mu\Gamma_\mu
	=: X_0 + \eta^\mu\Gamma_\mu\ ;
$$
the $a^i$ are determined from the equations of motion and the
stabilization algorithm; $\eta^\mu$ ($\mu=1,\cdots,P_1$) are
arbitrary functions of time; and
$$	\Gamma_\mu = {}^{(1)}\gamma^i_\mu {\partial\over\partial\dot q^i}\ ,
$$
where ${}^{(1)}\gamma^i_\mu$ are a subset of the null vectors of
$W_{ij}$, corresponding to the primary first class constraints found in
the Hamiltonian formalism.  It is not necessary to use the Hamiltonian
technique to find the $\Gamma_\mu$, but it does facilitate the
calculation:
$$	{}^{(1)}\gamma^i_\mu= {\partial\phi^1_\mu\over\partial p_i}(q,\hat p)\ ,
$$
where the $\phi^1_\mu$ are the primary first class constraints, and
$\hat p_i$ stands for the Lagrangian definition of the momenta
$\hat p_i=\partial L/\partial\dot q^i$.
The $P_1$ number of $\eta^\rho$ is the same number as in the
Hamiltonian formalism.  There are left $M-P_1$ constraints [3].

At this point it is useful to appeal to the Hamiltonian formalism for
the computation of the $P_1$ independent gauge transformations.  The
result includes the definitions of the functions $\phi^{j_\mu}_\mu$,
and then in the Lagrangian formalism we define
$$	f^i_{\mu,j_\mu}(q,\dot q) :=
	{\partial\phi^{j_\mu}_\mu\over\partial p_i}(q,\hat p)\ .
$$
These functions give the infinitesimal Lagrangian gauge transformations as
$$	\delta_\mu q^i = \sum^{K_\mu}_{j_\mu=1}
		\epsilon_\mu^{(K_\mu-j_\mu)} f^i_{\mu,j_\mu}\ ,
$$
the $\epsilon_\mu$ being arbitrary functions of time.

As we did in the Hamiltonian formalism, the first step
in the gauge fixing procedure will
be to fix the dynamics to determine
the arbitrary functions $\eta^{\mu}$. To this end we introduce $P_1$
constraints, $\omega^0_{\nu} \simeq 0$, such that
$D_{\mu\nu}:=\Gamma_{\mu}\omega^0_{\nu}$ has non-zero determinant:
$\det|\Gamma_{\mu}\omega^0_{\nu}| = \det|D_{\mu\nu}| \neq 0$.
Then the functions $\eta_{\mu}$ become determined by requiring the stability
of these new constraints:
	$$	0 = X\omega^0_{\nu}
		= X_0\omega^0_{\nu} + \eta^{\mu}\Gamma_{\mu}\omega^0_{\nu}\ .
	$$
This relation gives
	$$	\eta^{\mu} = -(D^{-1})^{\nu\mu}(X_0\omega^0_{\nu})\ ,
	$$
which determines the dynamics as
	$$	X_F = X_0 -(X_0\omega^0_{\nu})(D^{-1})^{\nu\mu}\Gamma_{\mu}\ .
	$$

Although the time evolution is fixed, as in the previous section
there still remain some point gauge transformations
in the constraint surface that we should get rid of.
Again, these transformations may be thought as affecting the
space of initial conditions.  In fact, we can extract
those transformations at $t=0$ that preserve the gauge fixing constraints
$\omega^0_{\nu}\simeq0$ from the general gauge transformations.  This
general transformation is
	$$	\delta\omega^0_{\nu} =\sum_{\mu=0}^{P_1}\delta_{\mu}\omega^0_{\nu}
		= \sum_{\mu=0}^{P_1}
			({\partial\omega_{\nu}^0\over\partial q^i}\delta_{\mu}q^i
			+ {\partial\omega_{\nu}^0\over\partial\dot{q}^i}
				\delta_{\mu}\dot{q}^i )\ ,
	$$
where now
	$$	\delta\dot{q}^i = {d\over dt}\delta {q}^i
		= X_F  \delta q^i + {\partial\delta q^i\over\partial t}\ .
	$$
In this expression we use the values at $t=0$:
	$$\eqalign{	\delta_\mu q^i(0)
			&=\sum_{i_\mu=1}^{K_\mu}
				\epsilon_\mu^{(K_\mu-i_\mu)}(0) f^i_{\mu,i_\mu}
			=\sum^{K_\mu}_{i_\mu=1}\alpha_{\mu,i_\mu}f^i_{\mu,i_\mu}\ , \cr
		{\partial\delta_{\mu}q^i\over\partial t}(0)
			&= \sum_{i_\mu=1}^{K_\mu}
				\epsilon_{\mu}^{(K_\mu-i_\mu+1)}(0) f^i_{\mu,i_\mu}
			= \sum_{i_\mu=0}^{K_\mu} \alpha_{\mu,i_\mu} f^i_{\mu,i_\mu+1}\ ,\cr
	}$$
after redefining $\alpha_{\mu,i_\mu} :=
\epsilon_\mu^{(K_\mu-i_\mu)}(0)$ (we have defined
$f^i_{\mu,K_{\mu}+1}=0$).

Notice that now, due to the presence of the time derivative of
$\delta_\mu q^i$,
$i_\mu$ runs from $0$ to $K_\mu $.
This is a key difference with respect to the Hamiltonian case, where
the $\alpha_{\mu,i_\mu}$ parameters had indices
$i_{\mu}$ running from $1$ to $K_\mu$.  We call the result for the
independent point gauge transformations (at $t=0$)
$\delta_{(0)}\omega^0_{\nu}$:
	$$	\delta_{(0)}\omega^0_{\nu}
		= \sum_{\mu=1}^{P_1} \biggl(\sum_{i_\mu= 1}^{K_\mu}
			\Bigl({\partial\omega^0_\nu\over\partial q^i}f^i_{\mu,i_\mu}
			+{\partial\omega^0_\nu\over\partial\dot{q}^i}(X_F f^i_{\mu,i_\mu})
			+ {\partial\omega^0_\nu\over\partial\dot{q}^i}
				f^i_{\mu,i_\mu+1}\Bigr)\alpha_{\mu,i_{\mu}}
			+\alpha_{\mu,0} \Gamma_\mu \omega^0_\nu\biggr)\ ,
	$$
where $f^i_{\mu,1}=\gamma^i_\mu$.
At this point, recalling that
$\det|\Gamma_\mu\omega^0_\nu| \neq 0$, we see that the stability conditions
$\delta_{(0)}\omega^0_\nu=0$ allow the determination of
$\alpha_{\mu,0}$ in terms of
$\alpha_{\mu,i_\mu}$ $(\mu=1,\cdots,P_1)$ $(i_\mu=1,\cdots,K_\mu)$.

We conclude that the independent point gauge transformations
$\delta_{(0)}$
that still remain, relating physically equivalent initial
conditions, are parameterized by
$\alpha_{\mu,i_\mu}$ $(\mu=1,\cdots,P_1)$ $(i_\mu=1,\cdots,K_\mu)$.
Their number equals $F$, the total number of first class
constraints in the Hamiltonian
theory. To eliminate these transformations we introduce
$F$ new gauge fixing constraints
$\omega^{i_\mu}_\mu\simeq0$ $(\mu=1,\cdots,P_1)$ $(i_\mu=1,\cdots,K_\mu)$,
with the conditions:
\item{1:}
	The system $\delta_{(0)} \omega^{i_\mu}_\mu =0$, which is
	linear in the $\alpha_{\mu,i_\mu}$ $(\mu=1,\cdots,P_1)$
	$(i_\mu=1,\cdots,K_\mu)$ has only the solution
	$\alpha_{\mu,i_\mu} = 0$ (so that no point gauge
	transformations are left).
\item{2:}
	$X_F(\omega^{i_\mu}_\mu) \simeq 0$
(the requirement of stability under evolution).

Now we have completed the gauge fixing procedure. For
reasons similar to the ones
raised in the Hamiltonian formalism, there are cases where a time dependent
constraint shows up necessarily.
Our examples in section 5 will be two of these cases, and we discuss
these cases in the next section.

Summing up, the gauge fixing constraints introduced in velocity space
(that is, in the Lagrangian formalism) are
$\omega^{i_\mu}_\mu$ $(\mu=1,\cdots,P_1)$ $(i_\mu=0,\cdots,K_\mu)$.
Their number is $F+P_1$, and therefore the total number of constraints
becomes $(M-P_1)+(F+P_1)=M+F$. The number of degrees of freedom is then
$2N-M-F$. Comparison with the results of the previous section shows
that we have proved
\proclaim Theorem.
The number of physical degrees of freedom in constrained Hamiltonian and
Lagrangian formalisms is the same.

Observe that this result, which was obviously expected on physical grounds,
is nontrivial.  In fact, before introducing the gauge fixing constraints, the
dimensions of the constraint surface were different in the two formalisms.
This means that the gauge fixing procedure has
to make up for this difference---and we see that it does.

\CC 4 Reparameterization Invariant Theories=

Reparameterization invariant theories provide interesting cases for the
application of the preceeding sections.  Examples of this kind,
including spatially homogeneous cosmologies of Types I and IX, will be
treated in the next section.

If we consider the infinitesimal
reparameterization $t\rightarrow t'=t-\epsilon(t)$, with $\epsilon$
an arbitrary (infinitesimal) function, the trajectories
$q^i(t)$ (any trajectory, not necessarily solutions of the equations of
motion) change accordingly, $q^i(t)\rightarrow q'^i(t')$. If we define the
functional infinitesimal transformation $\delta q^i=q'^i(t)-q^i(t)$, the
transformations we will consider are of the type
$\delta q^i=\epsilon\dot q^i+h^i$, where
the $h^i$ terms involve time derivatives of $\epsilon$.
The theory is reparameterization-invariant when the
Lagrangian remains form-invariant under these changes:
	$$	L(q(t),\dot{q}(t))dt = L(q'(t'),\dot q'(t'))dt'\ .
	$$
In such a case, we define
	$$	\delta L:={\partial L\over\partial q^i}\delta q^i +
		{\partial L\over\partial\dot q^i}\delta\dot q^i\ .
	$$
It is easy to check that $\delta L$ is a total derivative:
	$$	\delta L={d\over dt}(\epsilon L)\ .
	$$
This equality can be transformed into
	$$	[L]_i\delta q^i+ {d\over dt}G=0\ ,
	$$
where $[L]_i$ are the Euler-Lagrange derivatives and
where $G$ is the conserved quantity of Noether's theorem:
	$$	[L]_i={\partial L\over\partial q^i}\delta q^i +
			{\partial L\over\partial\dot q^i}\delta\dot q^i\ ,\
		G={\partial L\over\partial\dot{q}^i}\delta q^i - \epsilon L\ .
	$$

$G$ can be expanded as a sum of an $\epsilon$ term and terms involving
derivatives of $\epsilon$. Each of these terms
must be a constraint of the theory because $G$ is a constant of motion for
whatever arbitrary values we give to $\epsilon$
(this is a general argument for gauge symmetries).
Using the form of
of $\delta q^i$ introduced above, we obtain, for the $\epsilon$ coefficient
of $G$,
	$$	{\partial L\over\partial\dot{q}^i}\dot q^i - L=E_L\ ,
	$$
namely the Lagrangian energy function $E_L$.
Its corresponding canonical quantity
is the canonical Hamiltonian $H_C$; therefore we have stated the following:
\proclaim Theorem. The canonical Hamiltonian (if it is non-zero) in a
reparameterization-invariant theory is a constraint.

In general $H_C$ will be a secondary, first class constraint, but
a particular case is worth mentioning:  When all the configuration space
variables transform as $\delta q^i=\epsilon\dot q^i$ (``scalars''), then
$G$ becomes $G=\epsilon E_L$.  The relation
$[L]_i\delta q^i+{d\over dt}G=0$ gives
	$$	[L]_i\epsilon\dot q^i +\epsilon{d\over dt}E_L
			+ \dot\epsilon E_L = 0\ .
	$$
Since this relation is identically zero for any function $\epsilon$,
we conclude that the coefficient of $\dot\epsilon$, namely $E_L$, is
identically
zero.  Therefore the Lagrangian is homogeneous of first degree in the
velocities: The canonical Hamiltonian vanishes in this case.
The remaining pieces
tell us that $\dot q^i$ is a null vector of the Hessian matrix of the
Lagrangian, and in case this is the only null vector,
that there are no Lagrangian constraints. This situation occurs exactly
in the case of the
relativistic free particle, which is described by the Lagrangian
$L=\sqrt{\dot x^\mu\dot x_\mu}$ and which will be treated in the next
section.

Now we will prove another result for reparameterization-invariant theories,
the need for time dependence in some gauge fixing constraint.
Suppose that $H_C$ vanishes.  Then to fix the dynamics
(that is, to determine functions $\lambda^\mu$ in $H_D$ which do not
all vanish) by using the conditions $\dot\chi^1_\mu=0$, it is necessary
that at least one of the constraints $\chi^1_\mu$ have explicit time
dependence.  If $H_C$ doesn't vanish, then it is a constraint; the first
class Hamiltonian, $H_{FC}$, will necessarily be a first class constraint of
the original theory.  After the first step of the gauge fixing
procedure (in which the dynamics is determined), the final Hamiltonian $H_D$
will be a first class constraint that generates motions tangent to the
first-step gauge fixing surface.

This latter result means that $H_D$ will become a part of
$G(0)$, the generator of point gauge transformations which relate physically
equivalent initial conditions.  Then, in order to fulfill the two requirements
introduced in the second step of the gauge fixing procedure, it is
mandatory that at least one of the gauge fixing constraints be
time-dependent:  Otherwise there
is no way to satisfy the gauge fixing conditions.
By choosing variables appropriately, we can always end up with only one
time-dependent gauge fixing constraint.  Therefore we have proved:
\proclaim Theorem.  Reparameterization-invariant theories necessarily require
that one of the gauge fixing constraints be time-dependent.

This result is clearly expected from the physical interpretation
of this kind of theory:  The existence of reparameterization
invariance as a gauge symmetry implies that the evolution parameter
---the ``time''---is an unphysical variable.

\CC 5 Examples=

We will first discuss the case of a relativistic free particle and then
spatially homogeneous cosmological models of Bianchi Types I and IX.

One Lagrangian for a free particle in special relativity is ($\tau$ is
the path parameter)
	$$	L=\sqrt{\eta_{\mu\nu}\dot x^\mu\dot x^\nu}\ ,
	$$
where $\dot{}$ means $d/d\tau$.  (For convenience we take
$\eta_{\mu\nu}={\rm diag}(1,-1,-1,-1)$.)
The action integral $\int Ld\tau$ is invariant under arbitrary
reparameterizations, so $\tau$ is not necessarily proper time.
The conjugate momenta functions in velocity space are
	$$	\hat p_\mu={\eta_{\mu\nu}\dot x^\nu\over
			\sqrt{\eta_{\mu\nu}\dot x^\mu\dot x^\nu}}\ ,
	$$
and therefore the velocity-space energy function is
	$$	E_L=\hat p_\mu\dot x^\mu-L=0\ .
	$$

First, we examine this system from the Lagrangian point of view.
The definition of $\hat p_\mu$ implies
	$$	\eta^{\mu\nu}\hat p_\mu\hat p_\nu=1\ .
	$$
The equations of motion imply that $\hat p_\mu=$const.  The dynamics vector is
	$$	X= \dot x^\mu {\partial\over\partial x^\mu}
			+\lambda\bigl(\dot x^\mu {\partial\over\partial\dot x^\mu}\bigr)
			+{\partial\over\partial t}\ .
	$$
To fix the dynamics (to determine the arbitrary function $\lambda$), we
use the constraint that the path parameter is proper time:
	$$	\eta_{\mu\nu}\dot x^\mu\dot x^\nu=1\ ;
	$$
this constraint implies $\lambda=0$.

We must now set initial data, which will be seen to be equivalent to
fixing the zero point of proper time.  Because the Lagrangian is
homogeneous of first degree in the velocities, the gauge transformation
is of the form $\delta x^\mu=\epsilon\dot x^\mu$.  At $\tau=0$, this
becomes
	$$	\delta_0x^\mu=\alpha\dot x^\mu\ ,\
		\delta_0\dot x^\mu=\beta\dot x^\mu\ ,
	$$
where $\alpha,\beta$ are infinitesimal constants.  To be compatible with
the proper time constraint, we must have $\beta=0$.  We will choose
a constraint $\chi$ such that $\delta_0\chi=0$ implies $\alpha=0$
(to prevent any point gauge transformation); therefore
	$$	\dot x^\mu {\partial\chi\over\partial x^\mu}\neq 0\ .
	$$
The evolution of $\chi$ obeys
	$$	\dot x^\mu {\partial\chi\over\partial x^\mu} +
		{\partial\chi\over\partial\tau} =0\ ,
	$$
and so $\chi$ must be explicitly time-dependent.  One convenient choice is
	$$	\chi=x^0-\dot x^0\tau\ .
	$$
The zero point of the path
parameter is set by this requirement: $x^0(0)=0$.  There are then a
six-parameter set of paths which are solutions, the six parameters
being the three positions and three spatial components of the velocity
at $\tau=0$.

We now turn to the Hamiltonian discussion.
The Legendre map to phase space is a map onto the surface defined by
$\eta^{\mu\nu}p_\mu p_\nu=1$.  The canonical Hamiltonian $H_C$ is a
function on the surface, but a trivial one:  $H_C=0$.
The actual Hamiltonian in the Dirac procedure is the addition of an
appropriate function of the constraint to $H_C$, namely
	$$	H_D=\HALF\lambda(\tau)(\eta^{\mu\nu}p_\mu p_\nu-1)\ .
	$$
The equations of motion in phase space are $\dot p_\mu=0$ and
$\dot x^\mu=\lambda\eta^{\mu\nu} p_\nu$.  The one primary constraint ($M=1$)
	$$\phi:=\HALF(\eta^{\mu\nu}p_\mu p_\nu-1)
	$$
is  first class, and there are no
secondary constraints ($P_1=1$).  The gauge transformations are
generated by the one function
	$$	G=\epsilon(t)\phi\ .
	$$

The gauge fixing procedure has two steps.  The arbitrariness in the
equations of motion represented by $\lambda$ requires a constraint
function $\chi$ defined so that
	$$	\{\chi,\phi\}\neq0\ .
	$$
Clearly this function must be time dependent, or else the result of
requiring $\dot\chi=0$ will be $\lambda=0$.  One choice is
	$$	\chi=\tau-x^0\ \Longrightarrow\
			\lambda={1\over p_0}\ .
	$$
The Hamiltonian is now definite:
	$$	H={1\over2p_0}(\eta^{\mu\nu}p_\mu p_\nu-1)\ .
	$$
The second step is to consider $G(0)=\alpha\phi$, where
$\alpha=\epsilon(0)$, as a generator of point gauge transformations.
In order for the Poisson Bracket of $\chi$ with $G(0)$ to vanish,
clearly $\alpha$ must be zero, so that there are no further gauge
fixing steps to perform.

The result of gauge fixing in the Hamiltonian treatment of this free
particle is therefore that motion in phase space is within the surface
defined by $\eta^{\mu\nu}p_\mu p_\nu=1$ (a 7-dimensional space), and
the motion is given by
	$$	p_\mu={\rm const}\ ,\
		x^\mu={\eta^{\mu\nu}p_\nu\over p_0}\tau +{\rm const}\ .
	$$
Note that there are six free parameters for these paths, since the
$\chi$ constraint fixes the $x^0$ motion to be $x^0=\tau$.  Of course,
this is the same number as found in the Lagrangian treatment, though
here it was convenient to normalize $\tau$ by the requirement $\dot
x^0=1$, rather than by the requirement that $\tau$ be proper time (to
which it is proportional, anyway).

Our next set of examples are spacetime metrics which are invariant under a
three-dimensional isometry group which is transitive on spacelike three
surfaces [15].  The metric is best expressed
in a basis of differential
forms which is invariant under the group.  One such basis consists of
the four one-forms $\{dt,\omega^i\}$, where
	$$	d\omega^i=\HALF C^i_{st}\omega^s\wedge\omega^t\ ,
	$$
where the $C^i_{st}$ are the structure constants of the Lie algebra of
the group and obey
	$$	C^i_{st}=-C^i_{ts}\ ,\ C^a_{s[i}C^s_{jk]}=0\ .
	$$
The second relation is the Jacobi identity, and in the case of a
three-dimensional Lie algebra is exactly equivalent to
	$$	C^t_{st}C^s_{ij}=0\ .
	$$
In what follows, Greek indices range over 0,1,2,3, while Latin indices
range over 1,2,3.  The summation convention will be followed even if
both indices are superscripts or subscripts.

In this basis, the line element is
	$$	ds^2=-N^2dt^2+g_{st}(N^sdt+\omega^s)(N^tdt+\omega^t)\ ,
	$$
where $N$ is the lapse function and $N^s$ is the shift vector; $N$,
$N^s$, and $g_{st}$ are functions only of the time $t$.
The models are classified according to a standard listing of the
possible structure coefficients into nine basic classes, called Bianchi
Types.

Our examples will be the models of Type~I ($C^i_{jk}=0$) and Type~IX
($C^i_{jk}=k\epsilon_{ijk}$, $\epsilon_{ijk}$ is the completely
antisymmetric Levi-Civita symbol defined by $\epsilon_{123}=1$).  The
constant $k$ in the latter models is
redundant but is included to allow the limit $k=0$.  In both cases the
shift vector may be taken to be zero and the spatial metric $g_{st}$
may be taken to be diagonal (we treat only vacuum models).  In both
cases (as well as in any model which has $C^s_{si}=0$), the Lagrangian
of the system may be calculated from the spatially homogeneous form of
the scalar curvature $R$.

Instead of the above basis, it is more convenient to use an orthonormal
basis $\{\sigma^\mu\}$.  Since we take $N^i=0$ and $g_{ij}$ diagonal,
this basis is defined by
	$$	\sigma^0=Ndt\ ,
		\sigma^i=e^{-\Omega}e^{\beta_i}\omega^i\ ({\rm no\ sum\ on}\ i)\ ,
	$$
where $\Omega$ and $\beta_i$ are functions only of $t$, with $\sum\beta_i=0$.
The metric components $g_{ij}$ are given by
	$$	 (g_{ij})={\rm diag}(e^{-2\Omega+2\beta_i})\ .
	$$
In this basis the line element is
	$$	ds^2=\eta_{\mu\nu}\sigma^\mu\sigma^\nu\ ,\
			\eta_{\mu\nu}={\rm diag}(-1,1,1,1)\ .
	$$
Since $\sum\beta_i=0$, we define $\beta_\pm$ by
	$$	\beta_1 =\beta_++\ST\beta_-\ ,\
		\beta_2 =\beta_+-\ST\beta_-\ ,\
		\beta_3 =-2\beta_+\ .\
	$$

It is also possible to reparameterize time to make $N=1$ (or some
other function), but it will be seen that $N$ must remain a dynamical
variable, at least at first, in order not to spoil the Lagrangian
procedure.

The field equations for the functions $N,\Omega,\beta_i$ are
the Einstein equations (for a vacuum).  The
Ricci tensor coefficients $R_{\mu\nu}$ are functions only of $t$ in this
case, so the equations are ordinary differential equations.  The most
convenient form of these equations for our purposes will involve the
Einstein tensor $G_{\mu\nu}=R_{\mu\nu}-\HALF R\eta_{\mu\nu}$.  (The
equations may be written either as $G_{\mu\nu}=0$ or $R_{\mu\nu}=0$.)

Our first example is Bianchi Type I, in which $C^i_{jk}=0$.  In this
case the one-forms $\omega^i$ are expressible in terms of coordinates as
$\omega^i=dx^i$.  The line element is
	$$	ds^2=\eta_{\mu\nu}\sigma^\mu\sigma^\nu\ ,\
		\sigma^0=Ndt\ ,\
		\sigma^i=e^{-\Omega}e^{\beta_i}dx^i\ ({\rm no\ sum\ on}\ i)\ .
	$$
The appropriate Einstein tensor components (or rather, independent
linear combinations of them), which are to be set equal to zero, are
	$$\eqalign{{N^2\over3}G_{00}&
			=-\dot\Omega^2 +\dot\beta_+^2+\dot\beta_-^2=0\ ,\cr
		{N^2\over6}(G_{11}+G_{22}-2G_{33})&
			= \ddot\beta_+ -{\dot N\over N}\dot\beta_+
				-3\dot\beta_+\dot\Omega=0\ ,\cr
		{N^2\over2\ST}(G_{11}-G_{22})&
			= \ddot\beta_- -{\dot N\over N}\dot\beta_-
				-3\dot\beta_-\dot\Omega=0\ ,\cr
		{N^2\over6}(G_{11}+G_{22}+G_{33}+3G_{00})&
			=\ddot\Omega -{\dot N\over N}\dot\Omega
				-3\dot\Omega^2 =0\ .\cr
	}$$

The action integral for general relativity is ${\cal I}=\int R\sqrt{|g|}d^4x$
(up to an irrelevant proportionality constant), where $R$ is the
Ricci scalar and $g$ is the determinant of the metric
in coordinates $\{x^\mu\}$.  In our case we have
	$$	{\cal I} =\int R\; \sigma^0\wedge\sigma^1\wedge\sigma^2\wedge\sigma^3
			=\int RNe^{-3\Omega}d^4x\ .
	$$
The result here is
	$$	{\cal I} =\int {6e^{-3\Omega}\over N}(-\ddot\Omega
			+{\dot N\over N}\dot\Omega + 2\dot\Omega^2
			+\dot\beta_+^2 +\dot\beta_-^2)d^4x\ .
	$$
We take the volume of space $\int d^3x=1/12$ and integrate by parts with
respect to time, dropping the endpoint contributions:
	$$	{\cal I} =\int {e^{-3\Omega}\over 2N}
			(-\dot\Omega^2+\dot\beta_+^2 +\dot\beta_-^2)dt\ .
	$$
Variations of $\cal I$ with respect to $N,\beta_+,\beta_-,\Omega$
give the Einstein equations listed above.

Note that reparameterization of time can be used to set $N=1$ (or some
other function), but if $N$ is eliminated from the Lagrangian, the
$G_{00}=0$ field equation (which is a constraint equation) will not be
derivable.

But of course the Lagrangian $L$,
	$$	L={e^{-3\Omega}\over 2N}
			(-\dot\Omega^2+\dot\beta_+^2 +\dot\beta_-^2)\ ,
	$$
is singular, since $\dot N$ doesn't appear:
	$$	\hat p_N={\partial L\over\partial\dot N}=0\ .
	$$
An attempt to form the Hamiltonian thus yields
	$$	H_C=\HALF Ne^{3\Omega}(-p_\Omega^2+p_+^2+p_-^2)\ .
	$$
$H_C$ is a function in $T^*\!Q$, namely in principle a function of
$N,\Omega,\beta_+,\beta_-$, and $p_N,p_\Omega,p_+,p_-$, but it happens
to be independent of $p_N$.  The Legendre map from $TQ$ to $T^*\!Q$ is
	$$\{N,\Omega,\beta_+,\beta_-\}
			= \{N,\Omega,\beta_+,\beta_-\}\ ,
	$$
	$$	p_N=0\ ,
		p_\Omega=-{e^{-3\Omega}\over N}\dot\Omega\ ,
		p_+={e^{-3\Omega}\over N}\dot\beta_+\ ,
		p_-={e^{-3\Omega}\over N}\dot\beta_-\ ,
	$$
and thus maps the 8-dimensional $TQ$ into the 7-dimensional subspace of
$T^*\!Q$ defined by $p_N=0$.  We thus identify the primary constraint
in $T^*\!Q$:
	$$	p_N\simeq0\ ,
	$$
where $\simeq0$ means that every solution of the equations of motion has
to satisfy the constraint.  Since there is only the one primary
constraint, it is first class.

Therefore, to $H_C$ may be added an arbitrary function of
$p_N$ which vanishes when $p_N=0$, the simplest being $p_N$ itself:
	$$	H=H_C+\lambda p_N\ .
	$$
This arbitrariness means that we have a gauge-type freedom.
The time derivative of $p_N$ is given by
	$$	\dot p_N=\{p_N,H\}=-{\partial H\over\partial N}
			=-\HALF e^{3\Omega}(-p_\Omega^2+p_+^2+p_-^2)\ .
	$$
The requirement that $\dot p_N\simeq0$ thus implies that
$H_C/N\simeq0$ or $-p_\Omega^2+p_+^2+p_-^2\simeq0$.

Now is the time to generalize this example, so as not to repeat a lot of
material.  The generalization will still be quite concrete:  The Type IX
cosmology has
	$$	C^i_{jk}=k\epsilon_{ijk}\ .
	$$
In this case, the action integral is
	$$	{\cal I} =\int R\;
			\sigma^0\wedge\sigma^1\wedge\sigma^2\wedge\sigma^3\ ,
	$$
where the orthonormal frame of one-forms is
	$$	\sigma^0=Ndt\ ;\ \sigma^i
			=e^{-\omega}e^{\beta_i}\Omega^i\ ({\rm no\ sum\ on}\ i)\ ;\
				d\omega^i=\HALF k\epsilon_{ijk}\omega^j\wedge\omega^k\ .
	$$
Thus $\cal I$ is
	$$	{\cal I} =\int RNe^{-3\Omega}\;
			dt\wedge\omega^1\wedge\omega^2\wedge\omega^3\ ,
	$$
and the spatial integral may be set equal to 1/12 as before.

The result for the Lagrangian is ($\beta_+,\beta_-$ are defined as before):
	$$	L={e^{-3\Omega}\over 2N}
		(-\dot\Omega^2+\dot\beta_+^2 +\dot\beta_-^2)
			-k^2Ne^{-\Omega}V(\beta_+,\beta_-)\ ,
	$$
where the function $V$ is
	$$	V(\beta_+,\beta_-)= \HALF e^{-8\beta_+}
			-2e^{-2\beta_+}\cosh(2\ST\beta_-)
			+e^{4\beta_+}(\cosh(4\ST\beta_- -1)\ .
	$$
The particular form of $V$ is not important, nor are the specific forms
of the field equations.

Needless to say L is still singular:
	$$	\hat p_N=0\ .
	$$
There is still just one primary, first class constraint, $p_N\simeq0$, and the
Hamiltonian is given by
	$$	H=H_C+\lambda p_N\ ,
	$$
where
	$$	H_C=\HALF Ne^{3\Omega}(-p_\Omega^2+p_+^2+p_-^2)
			+k^2Ne^{-\Omega}V(\beta_+,\beta_-)\ .
	$$
Stability of the primary constraint gives
	$$	\dot p_N =\{p_N,H_C\} =-\psi \simeq 0\ ,
	$$
where $\psi$ is the secondary constraint
	$$	\psi = \HALF e^{3\Omega}(p_+^2 + p_-^2 - p_{\Omega}^2)
		+k^2e^{-\Omega} V(\beta_+,\beta_-)\ .
	$$
At this point we can see that the canonical Hamiltonian is just a
constraint: $H_C = N\psi$. No further constraints appear.

Primary and secondary constraints are both first class, the gauge
generator is made out of them:
	$$	G =\dot{\epsilon}p_N +\epsilon\psi\ .
	$$
After the redefinition of the arbitrary function,
$\epsilon=N\eta$, the transformations are
	$$\delta\beta_+=\dot{\beta}_+\eta\ ,
		\delta\beta_-=\dot{\beta}_-\eta\ ,
		\delta\Omega=\dot{\Omega}\eta\ ,
		\delta N=\dot{N}\eta +N\dot\eta\ .
	$$
(Thus, $\beta_+,\beta_-,\Omega$ behave as
scalars under reparameterizations, and $N$ behaves as a vector.)

According to the theory developed in the previous sections, there
will be two gauge fixing constraints, one of them time-dependent. These
two constraints have to make the two original constraints second class in the
theory in order to determine the arbitrary function in the Dirac Hamiltonian.
The simplest way to proceed is
to reverse our formal methodology and
first to write down the initial data gauge fixing constraint.
If this constraint does not depend on $N$, its stability will
give us a new constraint which is $N$-dependent. Then the stabilization of
this new constraint will fix the dynamics.

We start with the time-dependent constraint (a simple and common choice):
	$$	\chi^{(1)} :=\Omega-t \simeq0\ .
	$$
Its stability leads to the requirement
	$$	\dot\chi^{(1)} = \{\Omega,H_C\} -1  =-Ne^{-3\Omega}p_\Omega -1
		\simeq 0\ .
	$$
 From this expression we get the new gauge fixing constraint
	$$	\chi^{(2)} := N +{e^{3\Omega}\over p_\Omega} \simeq 0\ .
	$$
Notice that since $N$ is required to be positive (on physical grounds),
this constraint implies $p_\Omega<0$.

In its turn, stability of $\chi^{(2)}$, because of the presence of $N$, will
determine the arbitrary function $\lambda$ in the Dirac Hamiltonian
$H_D=H_C+\lambda p_N$.  Actually, in order to fix the gauge in the
equations of motion, we do not need to know the value of $\lambda$:
It only matters for the equations of motion for the
variable $N$, and the constraint
$\chi^{(2)}\simeq0$ already relates $N$ to other variables.
Therefore, since
we are concerned with the evolution of $\beta_+,\beta_-,\Omega$ and
their associated momenta, the Hamiltonian is simply:
	$$	H = -{1\over2p_\Omega}(p_+^2+p_-^2-p_\Omega^2
			+2k^2e^{-4\Omega}V(\beta_+,\beta_-))\ ,
	$$
where we have used the constraint $\chi^(2)\simeq0$ to substitute for
$N$. This procedure is correct
because $N$ was an overall factor of a constraint in $H_C$: $H_C=N\psi$.
This result can be rephrased as follows:
We have gotten rid of a couple of canonical
variables, $N,p_N$, by using the Dirac Bracket. In this case, the Dirac
Bracket for the rest of the variables is the usual Poisson Bracket..

The constraint $\psi\simeq0$ can be used to define the evolution of $p_\Omega$
in terms of the other variables (since the constraint $\Omega=t$ fixes the time
dependence of $\Omega$).  It is conveniently factorized as
	$$	\psi=-\HALF e^{3\Omega}(p_\Omega+H_R)(p_\Omega-H_R)\ ,
	$$
where
	$$	H_R :=\sqrt{p_+^2+p_-^2 +2k^2e^{-4\Omega}V(\beta_+,\beta_-)}\ .
	$$
This factorization implies
	$$	H = {1\over2p_\Omega}(p_\Omega-H_R) (p_\Omega+H_R)\ .
	$$

Satisfaction of the constraint $\psi\simeq0$ implies
$ p_\Omega+H_R\simeq0$ (the physical interpretation of the model
requires $p_\Omega\leq0$). Then
$p_\Omega-H_R\simeq2p_\Omega $, and the Hamiltonian can be
equivalently written as
	$$	H =p_\Omega+H_R\ .
	$$
Observe that this constraint defines the evolution
of $p_\Omega$ in terms of the other variables.
Thus if we restrict ourselves to the evolution of the variables
$\beta_+,\beta_-$ and their canonical conjugates $p_+,p_-$, then the
dynamics in this reduced space is described by the Hamiltonian $H_R$.
Once the evolution of these variables is determined, the rest of
the variables (four in number) have a time-evolution dictated by the
constraints (also four in number) of the theory.

With regard to the Lagrangian formulation, there is only one constraint in
velocity space:  It is the pullback of the secondary constraint $\psi$,
since the pullback of the primary one is identically zero.  The dynamics
(given by a vector field in
velocity space) has only one arbitrary function, which multiplies
the vector field $\Gamma=\partial/\partial\dot N$.

Let us briefly sketch the Lagrangian gauge fixing procedure in this case.
If we start again with the time dependent gauge fixing constraint
$\Omega-t\simeq0$, its stabilization
will require $\dot\Omega-1\simeq0$.  The stabilization of
this new gauge fixing constraint will give a new constraint which is
dependent on $\dot N$ (plus other velocity space variables).  Finally,
the stabilization
of this last gauge fixing constraint determines the arbitrary function
of the dynamics.  We end up with three gauge fixing constraints that add to the
original single
constraint of the theory to give the elimination of four degrees of
freedom.  This numbering is the same as in the Hamiltonian case.

\CC 6 Conclusions=

A singular Lagrangian, that is with singular Hessian
$W_{ij}=\partial^2L/\partial\dot q^i\partial\dot q^j$,
results in constraints, a dynamics with some arbitrariness, and gauge
transformations which reflect this arbitrariness.  The Legendre map to
phase space defined by $p_i=\partial L/\partial\dot q^i$
therefore maps the $2N$ dimensional velocity space $TQ$ to a lower
dimensional surface in phase space $T^*\!Q$.
The requirement that the dynamics on this surface be consistent can be
used to reduce its dimensionality somewhat, but there
still remain the same three ingredients:  constraints, arbitrariness in
the dynamics, and gauge transformations.
Consistency requirements in the Lagrangian formalism can also be used to
reduce somewhat the dimensionality of the constraint surface in
velocity space, but in general the constraint surface in phase space
will have smaller dimensionality than the constraint surface in
velocity space.  The number of gauge transformations in $TQ$ and in
$T^*\!Q$ is the same.

In this paper we show how to determine the dynamics and fix the gauge
in both the Hamiltonian and Lagrangian formalisms.  We believe the
Lagrangian discussion is new and useful.  The result includes a proof
that the final number of degrees of freedom in the two formalisms is
the same.
Important parts of our methods are the two steps of determining the
dynamics and determining the independent initial data.

The first step in both the Lagrangian and Hamiltonian formalisms is the
stabilization algorithm (which ensures consistent dynamics).  The
result is $M$ constraints in the Hamiltonian case (a constraint surface
of dimension $2N-$M in $T^*\!Q$)
and $M-P_1$ constraints in the Lagrangian case (a constraint surface of
dimension $2N-M+P_1$ in $TQ$);
$P_1$ is the number of primary, first class constraints.

There are $P_1$ gauge transformations (in both the Lagrangian and
Hamiltonian cases), and they are formed using a total of $F$ functions
(first class constraints in the Hamiltonian formalism).  In the second
step, in both the Lagrangian and Hamiltonian cases, $P_1$ gauge-fixing
functions are used to determine the dynamics (to obtain the dynamics
vector $X_F$ or the Hamiltonian $H_D$ respectively).  The
dimensionality of the constraint surfaces in each case is reduced by
$P_1$.

The third and final step is to use gauge fixing functions to determine
physically inequivalent initial data.  In the Hamiltonian case $F-P_1$
functions are used, so that the number of degrees of freedom is
$2N-M-P_1-(F-P_1)=2N-M-F$.  In the Lagrangian case $F$ functions are
used, and again the number of degrees of freedom is
$2N-M+P_1-P_1-F=2N-M-F$.  The Hamiltonian method directly shows that
$2N-M-F$ is an even number.

We have emphasized the double role played by the gauge fixing
procedure.  Some constraints are needed to fix the time
evolution of the gauge system, which was undetermined to a certain
extent. The rest of the constraints eliminate the spurious degrees of
freedom that are still present in the setting of the initial conditions
of the system. This double role has not been adequately emphasized in
the literature.
Although we only treat classical systems in this paper, our approach
should also be relevant to quantum ones.

Let us briefly comment on Dirac's extended Hamiltonian formulation
[2].  Dirac suggests that the canonical dynamics be modified by adding
all secondary first class constraints to
the Hamiltonian in an {\it{}ad hoc\/} manner.
The result is as many arbitrary functions in the
dynamics as first class constraints, and every first class constraint
generates an independent gauge transformation.
In our dynamics fixing step, we would introduce F constraints, and the
process would then be finished (the initial data fixing step would be
empty).  The counting of degrees of freedom still agrees with ours.
Furthermore, if we take one of our sets of gauge fixing constraints
$\chi^{i_\mu}_\mu\simeq0$, then this set works for the extended
formalism also, and we end up with the same dynamics.  However, one can
also use a more general gauge fixing procedure for the extended
Hamiltonian theory (without the stabilization condition we required
for the $\chi^{i_\mu}_\mu$); in
this case the extended Hamiltonian theory will not necessarily be
equivalent to the Lagrangian formalism.

We have discussed, also, the special case of
reparameterization-independent theories.  In those cases, as we showed,
the canonical Hamiltonian $H_C$ is a constraint of the motion.  In
general $H_C$ will be a secondary, first class constraint, but in some cases
it will vanish.  In reparameterization-independent theories, we showed
that one of the gauge fixing constraints must be time-dependent.
In these theories, the ``time'' is defined
through some dynamical variables of the system.  Since there is a good
deal of choice for this definition of time, one has to be very careful
in verifying that the time-dependent constraint is consistent with the
physical interpretation of the model.

We have illustrated these ideas by considering two interesting cases.
The first is the motion of a free particle in special relativity using
the Lagrangian $L=\sqrt{\eta_{\mu\nu}\dot x^\mu\dot x^\nu}$.  This
Lagrangian is reparameterization-invariant, and the canonical
Hamiltonian vanishes.  The result of applying our procedures is that
the paths are parameterized by six parameters (three positions in space
and three velocity or three momentum initial values), and that time may
be parameterized by proper time, as one does expect.

The second case included two spatially homogeneous cosmological models
in general relativity, the vacuum Bianchi Type I and Type IX models.
In these models, the number of degrees of freedom is found to be
reducible to four, and the role of the time-dependent gauge fixing
procedure is clarified.

We have previously [16] described some of the above results and intend
to extend many of these ideas, for instance by looking
into aspects of field theory.  For example, the various gauges
used in electromagnetic theory, including the Coulomb
($\vec\nabla\cdot\vec A=0$), Lorentz ($A^{,\sigma}\!_\sigma=0$), and radiation
($A_0=0$) gauges, apply in different formulations, either Lagrangian or
Hamiltonian.  How they affect the true degrees of freedom of the
electromagnetic field may be clarified by our methods.

\CC Acknowledgements=

J.M.P.\ acknowledges support by the Comisi\'on Interministerial para
la Ciencia y la Tecnolog\'\i{}a (project number AEN-0695) and by a
Human Capital and Mobility Grant (ERB4050PL930544).  He also thanks
the Center for Relativity at The University of Texas at Austin for
its hospitality.

\vfill\eject

\CC References=

\parindent=0pt
[1]
Dirac P A M 1950
{\it{}Can.\ J.\ Math.\/} {\bf{}2}, 129

[2]
Dirac P A M 1964
{\it{}Lectures on Quantum Mecanics\/}
(New York: Yeshiva Univ.\ Press)

[3]
Pons J M 1988
{\it{}J.\ Phys.\ A: Math.\ Gen.\/} {\bf{}21}, 2705

[4]
Gracia X and Pons J M 1988
{\it{}Annals of Physics (N.Y.)\/} {\bf{}187}, 2705

[5]
Batlle C, Gomis J, Pons J M, and Roman N 1986
{\it{}J.\ Math.\ Phys.\/} {\bf{}27}, 2953

[6]
Sundermeyer K 1982
``Constrained Dynamics'' in
{\it{}Lecture Notes in Physics\/} {\bf{}169} (Berlin: Springer Verlag)

[7]
Henneaux M, Teitelboim C, and Zanelli J 1990
{\it{}Nucl.\ Phys.\/} {\bf{}B332}, 169

[8]
Sudarshan E C G and Mukunda N 1974
{\it{}Classical Dynamics: A Modern Perspective\/} (New York: Wiley)

[9]
Anderson J L and  Bergmann P G 1951
{\it{}Phys.\ Rev.\/} {\bf{}83}, 1018

[10]
Castellani L 1982
{\it{}Annals of Physics (N.Y.)\/} {\bf{}143}, 357

[11]
Sugano R, Saito Y, and  Kimura T 1986
{\it{}Prog.\ Theor.\ Phys.\/} {\bf{}76}, 283

[12]
Gomis J, Henneaux M, and Pons J M 1990
{\it{}Class.\ Quantum Grav.\/} {\bf{}7}, 1089

[13]
Gotay M, Nester J M, and Hinds G 1978
{\it{}J.\ Math.\ Phys.\/} {\bf{}19}, 2388

[14]
Kamimura K 1982
{\it{}Nuovo Cimento\/} {\bf{}B69}, 33

[15]
Ryan M P and Shepley L C 1975
{\it{}Homogeneous Relativistic Cosmologies\/} (Princeton:  Princeton Univ.\
Press)

[16]
Pons J M and Shepley L C 1994
``Gauge Fixing in Constrained Systems'' to appear in
Charap J M, Ed, {\it{}Geometry of Constrained Dynamical Systems\/}
(Cambridge: Cambridge Univ.\ Press)

\bye